\documentclass[11pt]{article}

\usepackage{cite}
\usepackage{epsf}
\usepackage{amsmath}
\usepackage{amsfonts}
\usepackage{amssymb}
\usepackage{bm}
\usepackage{bbm}

\usepackage{graphicx}
\usepackage{epsfig}
\usepackage{latexsym}

\def\corresponds{{\lower.2ex\hbox{=}}{\rm\kern-.75em^\triangle}}
\def\succsim{\succ\kern-.9em_\sim\kern.3em}
\def\precsim{\prec\kern-1em_\sim\kern.3em}
\def\slantfrac#1#2{\kern1em^{#1}\kern-.3em/\kern-.1em_{#2}}
\def\lfrac#1#2{{}^{#1\!}\kern-.0em/_{#2}}

\def\buildrel#1\under#2{\mathrel{\mathop{\kern0pt #2}\limits_{#1}}}

\begin{document}

\bibliographystyle{myprsty}

\vspace*{1.0cm}
\begin{center}
\begin{tabular}{c}
\hline
\rule[-5mm]{0mm}{15mm}
{\Large \sf The Study of the Heisenberg-Euler Lagrangian}\\
{\Large \sf and Some of its Applications}\\[2ex]
\hline
\end{tabular}
\end{center}
\vspace{0.2cm}
\begin{center}
S. R. Valluri
\end{center}
\begin{center}
{\it Departments of Physics, Astronomy \& Applied Mathematics,\\
University of Western Ontario, London, ON, N6A 3K7, Canada}\\
{\tt valluri@uwo.ca}
\end{center}
\begin{center}
U. D. Jentschura
\end{center}
\begin{center}
{\it National Institute of Standards and Technology,\\ 
Mail Stop 8401, Gaithersburg, Maryland, 20899--8401, USA}\\
{\tt ulj@nist.gov}
\end{center}
\begin{center}
D. R. Lamm
\end{center}
\begin{center}
{\it Electro-Optics, Environment, and Materials Laboratory,\\
GTRI, Georgia Institute of Technology, Atlanta, GA, 30332, USA}\\
{\tt darrell.lamm@gtri.gatech.edu}
\end{center}
\vspace{0.3cm}
\begin{center}
\begin{minipage}{11.8cm}
{\underline{Abstract}}
The Heisenberg-Euler Lagrangian (HEL) is not only a topic of
fundamental interest, but also has a rich variety of diverse
applications in astrophysics, nonlinear optics and elementary
particle physics etc. We discuss the series representation of this
Lagrangian and a few of its applications in this study.
[In an appendix, we discuss issues related to the renormalization
---and the renormalization-group invariance--- of the 
HEL and its two-loop generalization.]
\end{minipage}
\end{center}

\newpage

\section{INTRODUCTION}

The effective one-loop Lagrangian density [1--11] of quantum
electrodynamics (QED) describes the nonlinear interaction of
electromagnetic fields due to a single closed electron loop. 
[It is also referred to as the QED effective action, or
the effective Lagrangian.]
One of the most compact and elegant ways to treat the symmetry
properties of the vacuum is by the method of the effective
Lagrangian \cite{12}.

This one-loop Lagrangian, often called the Heisenberg-Euler
Lagrangian (HEL), has been used to describe a variety of
electromagnetic processes. The real part of the Lagrangian can be
used to delineate such dispersive phenomena as photon propagation
in a magnetic field, second harmonic generation, photon splitting
in a magnetic field, and light scattering in a vacuum
[1,4--9,11,13--17].  The development of high intensity lasers,
with the consequent availability of powerful, coherent light
sources, can render possible the observation of delicate nonlinear
effects and also the nonlinearity of Maxwell's 
equations. Strong magnetic fields around pulsars and around magnetars, and
other astrophysical and laboratory situations might also warrant
accurate calculations based on the Lagrangian for astrophysical and other
applications. The imaginary part of the Lagrangian has been
applied to absorption processes such as electron-positron pair
creation \cite{4,5,8,9}.

While the integral representation of this Lagrangian as a function
of the electric and magnetic field is well known, further
analytical representations and numerical procedures for arbitrary
field strengths have recently been refined. In the special cases
when either the magnetic or electric field vanishes, the
Heisenberg-Euler Lagrangian may be expressed in terms of
elementary functions and integrals of the natural logarithm of the
generalized gamma function with real or complex arguments \cite{18,19}.
Elementary function series expansions for these gamma function
integrals and precise numerical values for the Lagrangian using
the expressions of Dittrich et al. were provided by Valluri et al.
\cite{18}. For the more general case when both an electric and magnetic
field are present, a numerically useful, analytical series
expression for the real part of the Lagrangian was derived and
this latter series expression involves elementary functions and
the sine, cosine, and exponential integrals, all of which are
easily calculated \cite{20}.

Motivations for studying non-linear generalizations of Maxwell's
equations in the vacuum are now quite different. Strong interest
in \textbf{one-loop corrections} to the classical Lagrangian in
Abelian (as well as non-Abelian) gauge theories has resurged in
order to learn about the structure of the vacuum when it is probed
by an external electromagnetic field. The problem of the existence
of a stable electron has been and continues to be an interesting issue,
not only relevant to pure electrodynamics but also to 
an unknown fundamental theory of matter and
its interactions (for an interesting view on this issue, see
\cite{12}).

In this paper we discuss some analytic calculations relevant for
the applications of the HEL. The paper is organized as
follows.  Section 2 provides the analytical definitions and
preliminaries. Section 3 presents the analytical results
indicating the existence of the higher harmonics for the real part
of the HEL, and the discussion of these results will conclude the
main body of the paper. Some aspects of the renormalization 
and the renormalization-group invariance of the action are
briefly discussed in an Appendix. The final section summarizes the
conclusions.

%
%
\section{REPRESENTATION OF THE QED EFFECTIVE\\
ACTION BY SPECIAL FUNCTIONS}

The renormalized (see also the Appendix) Heisenberg-Euler
Lagrangian (HEL) $\Delta \mathcal{L}$ is expressed as a
one-dimensional proper-time integral [1--4]:
\begin{eqnarray}
\Delta \mathcal{L} = -\frac{e^2}{8\pi^2} \,
\lim_{\epsilon,\eta \to 0^+} \, 
\int_{\eta}^{i\infty+\eta} \, 
\frac{\,ds}{s}\,
e^{-(m^2-i\epsilon)s} \left[ab \coth(eas) \cot(ebs) -
\frac{a^2-b^2}{3}-\frac{1}{(es)^2} \right]
\end{eqnarray}
\mbox{ and is a quantum correction to the Maxwellian Lagrangian}
\begin{eqnarray}
L_{cl} = L_0 = L_M = \frac{1}{2}(E^2 - B^2)\,.
\end{eqnarray}
$\Delta \mathcal{L}$ is a one-dimensional proper time integral
with proper-time parameter $s$ and can be \textbf{evaluated by
numerical quadrature}. A \textbf{convergent series expansion} with
(computational and conceptual advantages) can be derived in terms
of  \textbf{special functions}. There is a unified series expansion
encompassing both the real and the imaginary part~\cite{20}. We need
techniques for a \textbf{reliable numerical evaluation} which
may have a variety of applications. While the series expansion
might be regarded as a complete solution from a theoretical point 
of view, the expansion alone does not solve the physics associated
with the Lagrangian, and it does not provide {\em eo ipso} a 
general, numerically efficient algorithm for its evaluation.
The integral defining the HEL, which can be
calculated numerically by quadrature, has a slow convergence due
to oscillations of the integrand.

The QED effective Lagrangian can be expressed as a function of the
Lorentz invariants $\mathcal{F}$ and $\mathcal{G}$, given by
\begin{eqnarray}
\mathcal{F} = \frac{1}{4}F_{\mu\nu} F^{\mu\nu} = \frac{1}{2}
\left( \bm{B}^2 - \bm{E}^2 \right) = \frac{1}{2} \left(
a^2 - b^2 \right)\,,
\end{eqnarray}
\begin{eqnarray}
\mathcal{G} = \frac{1}{4}F_{\mu\nu} \ (*F)^{\mu\nu} = -\bm{E}
\cdot \bm{B} = \pm ab\,,
\end{eqnarray}
where $\bm{E}$ and $\bm{B}$ are the electric and magnetic
field strengths, $F_{\mu\nu}$ is the field-strength tensor and
$(*F)^{\mu\nu}$ denotes the dual field-strength $(*F)^{\mu\nu} =
(1/2)\epsilon^{\mu\nu\rho\sigma}\ F_{\rho\sigma}$. The quantities
$a$ and $b$ denote \emph{secular invariants},
\begin{eqnarray}
a = \sqrt{\sqrt{\mathcal{F}^2 + \mathcal{G}^2} + \mathcal{F}}\,, \\
b = \sqrt{\sqrt{\mathcal{F}^2 + \mathcal{G}^2} - \mathcal{F}}\,.
\end{eqnarray}
Secular invariants emerge naturally as eigenvalues of the
\textbf{field-strength tensor} fields.  We also note that in the
case $\mathcal{G}<0$, it is possible to choose a Lorentz frame in
which $E$ and $B$ are parallel.
\begin{eqnarray}
a = \left| \bm{B} \right|  \mbox{ and } 
b = \left| \bm{E} \right| \mbox{ if } 
\bm{B} \mbox{ is (anti-)parallel to } \bm{E}\,,
\end{eqnarray}
$a$ and $b$ are positive definite and this gives the condition:
\begin{eqnarray}
ab = \left| \bm{E} \cdot \bm{B} \right| > 0 \mbox{ for any
Lorentz frame and } \mathcal{G} \ne 0\,.
\end{eqnarray}
Also, the following notation is sometimes used:
\begin{eqnarray}
\mathcal{F} = -S, \hspace{20pt} \mathcal{G} = -P\,,
\end{eqnarray}
\begin{eqnarray}
\mathcal{L}_{\mbox{cl}} = -\mathcal{F} =
-\frac{1}{4}F_{\mu\nu}F^{\mu\nu} = \frac{1}{2} \left( \bm{E}^2
- \bm{B}^2 \right) = \frac{1}{2} \left( b^2 - a^2 \right)\,.
\end{eqnarray}
The correction $\Delta \mathcal{L}$ to the Maxwellian Lagrangian
$L_0$ can be written in terms of the secular invariants $a$ and
$b$~\cite{8,9},
\begin{eqnarray}
\Delta \mathcal{L} =  {\rm Re}\Delta \mathcal{L} + 
{\rm i}\,{\rm Im}\Delta\mathcal{L}\,.
\end{eqnarray}
\noindent the real part can be expressed as
\begin{eqnarray}
\mbox{{\rm Re}}\  \Delta \mathcal{L} = -\frac{e^2}{4\pi^3}ab
\sum_{n=1}^\infty \left[ a_n + d_n \right]\,,
\end{eqnarray}
\begin{eqnarray}
a_n = \frac{\coth(n\pi b/a)}{n} \left\lbrace \mbox{Ci} \left(
\frac{n\pi m^2}{ea} \right) \cos\left(\frac{n\pi m^2}{ea}\right) +
\mbox{si} \left( \frac{n\pi m^2}{ea} \right) \sin \left(
\frac{m\pi m^2}{ea} \right) \right\rbrace ,
\end{eqnarray}
\begin{eqnarray}
d_n = \frac{-\coth(n\pi a/b)}{2n} &\Bigg\{&\exp \left(
\frac{n\pi m^2}{eb} \right) 
\mbox{Ei} \left(-\frac{n\pi m^2}{eb}\right)\nonumber \\&+& 
\exp \left( -\frac{n\pi m^2}{eb} \right)
\mbox{Ei} \left( \frac{m\pi m^2}{eb} \right) \Bigg\} ,
\end{eqnarray}
\noindent while the imaginary part is given as
\begin{eqnarray}
{\rm Im}\ \Delta \mathcal{L} = 
\frac{e^2\left|ab\right|}{8\pi^2}
\sum_{n=1}^{\infty} 
\frac{1}{n} 
\coth \left(\frac{n\pi a}{b}\right) \,
\exp\left(-\frac{n\pi m^2}{eb}\right)\,.
\end{eqnarray}
\noindent We note that the Cosine and Sine integrals have an
``asymmetric'' form since the generally accepted definitions for these integrals
are ``asymmetric'' and are shown below. More details are given in
\cite{20}.
\begin{eqnarray}
\mbox{Ci}(z) = -\int_{z}^{\infty} \, dt \ \frac{\cos(t)}{t}
\hspace{20pt} z > 0\,,
\end{eqnarray}
\begin{eqnarray}
\mbox{si}(z) = -\int_{z}^{\infty} \, dt \ \frac{\sin(t)}{t} =
\mbox{Si}(z) - \frac{\pi}{2}\,,
\end{eqnarray}
\begin{eqnarray}
\mbox{Si}(z) = \int_{0}^{z} \, dt \ \frac{\sin(t)}{t}\,.
\end{eqnarray}
\noindent The imaginary part Im$\Delta\mathcal{L}$ is generated
by a modification of the integration contour in the exponential
integral entering into the definition of $d_n$ (normally, the 
exponential integral is defined via a principal-value prescription).
The relevant exponential integral is
\begin{eqnarray}
\mbox{Ei} \left(\frac{n\pi m^2}{eb} \right) 
\mbox{in the definition of }\,\, d_n \,\,\mbox{and reads}\\
\mbox{Ei}(u) = - ({\rm P.V.}) \, \int_{-u}^{\infty} \frac{e^{-t}}{t} \, dt
\hspace{30pt}  \mbox{for $u \in \mathbbm{R}$. }
\end{eqnarray}
Under an appropriate deformation of the contour,
a unified representation for both the real and the imaginary
parts is obtained~\cite{20},
\begin{eqnarray}
\Delta \mathcal{L} = \lim_{\epsilon \to 0^+} -\frac{e^2}{4\pi ^3}
ab \sum_{n = 1}^{\infty} \left[b_n + c_n\right]\,,
\end{eqnarray}
\begin{eqnarray}
b_n = -\frac{\coth\left(n\pi b/a\right)}{2n} &\Bigg\{&\exp
\left(-i \frac{n\pi m^2}{ea} \right) \Gamma\left(0, -i \frac{n\pi
m^2}{ea}\right)\nonumber \\ &+& \exp \left(i \frac{n\pi m^2}{ea} \right)
\Gamma\left(0, i \frac{n\pi m^2}{ea}\right)\Bigg\}\,,
\end{eqnarray}
\begin{eqnarray}
c_n = \frac{\coth\left(n\pi a/b\right)}{2n} &\Bigg\{&\exp
\left(\frac{n\pi m^2}{eb} \right) 
\Gamma\left(0, \frac{n\pi m^2}{eb}\right)\nonumber \\ &+& 
\exp \left(-\frac{n\pi m^2}{eb} \right)
\Gamma\left(0, -\frac{n\pi m^2}{eb} + i\epsilon\right) \Bigg\}\,.
\end{eqnarray}
The effective action has branch cuts along the positive and 
negative $b$-axis as
well as along the positive and negative imaginary axis. For more details, the
reader is referred to~\cite{20}. In
contrast to the exponential integral Ei, the incomplete Gamma
function is defined in the entire complex plane with a cut along
the negative real axis. It is important to see this connection
since the Barnes function \cite{21} is closely connected to the Gamma
function which has numerous applications.

The main numerical difficulty is the slow overall convergence of
the series expansion whose terms are of nonaltering sign. 
Pad\'{e} approximants, a standard tool
in many power series application are not capable of summing the
series for the HEL \cite{20}.  The terms of the \textbf{convergent
series representation} are interpreted as being generated by a
\textbf{``partial-fraction decomposition''} in close analogy to
\textbf{``partial-wave decomposition''} in bound state
calculations. It has been shown that the convergence of the HEL
series can be accelerated by the same technique as in partial wave
decomposition also called the ``combined nonlinear-condensation
transformation'' (CNCT) \cite{20,22,23}.

Another interesting representation is in terms of the Barnes
function.  The Barnes function G(z) is a generalization of the
Euler gamma function and is also related to the Hurwitz Zeta
function. It obeys the recursion relation~\cite{21}
\begin{eqnarray}
\mbox{G}(z+1) = \Gamma(z) \, \mbox{G}(z) \hspace{20pt} z \in
\mathbbm{C}, \hspace{20pt} \mbox{G}(1) = 1\,.
\end{eqnarray}
We write a useful expression for $\log G(z+1)$ below:
\begin{eqnarray}
\log \, G(z+1) &=& \frac{1}{2} z \, \log(2\pi) \,\,
-\frac{\gamma z^2}{2} \,- \frac{z(z+1)}{2} \nonumber \\
&+& \sum_{k=2}^{\infty} (-1)^k \, \zeta(k) \frac{z^{k+1}}{k+1}\,,
\end{eqnarray}
\begin{eqnarray}
\log \mbox{G}(z+1) &=& z\log( \, \Gamma(z)) + \zeta'(-1) - \zeta'(-1, z)\,, \\
\zeta'(t, z) &=& \frac{d}{dt} \, \zeta(t, z)\,,
\end{eqnarray}
where $\zeta(k)$ is the Riemann Zeta function, used in the
representation of the HEL. Due to its close connection to the
$\Gamma(z)$ and the Zeta function, the Barnes function may have a
variety of applications in Computer Algebra and Theoretical
Physics as well as in other fields.

The Mittag--Leffler theorem (see~\cite{20} for a comprehensive
discussion) is a key element in the 
derivation of the series representation for the HEL
(some useful formulas, originally derived without the 
explicit use of the Mittag--Leffler theorem, still based on 
a partial-fraction decomposition and equivalent to the 
results obtained by us using the Mittag--Leffler
theorem, can be found in~\cite[p.~271]{24}).
The generalization of the Mittag-Leffler theorem for a function
with nonsimple poles and zeros  is the Riemann-Roch theorem
\cite{25,26}. For Lagrangians with poles and zeros of higher order,
which can occur in string theories, we anticipate the application
of the Riemann-Roch theorem. Such Lagrangians with poles and zeros
of arbitrary order may form a vector space of meromorphic
functions over a complex field.

%
%
\section{MAGNETO--OPTICAL EFFECT: SECOND\\
AND HIGHER--HARMONIC GENERATION}

Maxwell's equations receive corrections from virtual excitations
of the charged quantum fields(notably electrons and positrons).
This leads to interesting effects \cite{1}: light-by-light scattering,
photon splitting, modification of the speed of light in the
presence of strong electromagnetic fields, and -- last, but not
least -- pair production. The dominant effect for electromagnetic
fields that vary slowly with respect to the Compton wavelength
(frequencies $\omega \ll 2mc^2/h$) is described by the
Heisenberg-Euler Lagrangian, which is known to all orders in the
electromagnetic field. For the case of zero electric field, the
HEL can be written as \cite{18,19}
\begin{eqnarray}
\Delta \mathcal{L}(h) =
\frac{m^4}{32\pi^2}\Bigg[\left(\frac{1}{h^2}\right)
\bigg\lbrace-\left(\frac{1}{3}+2h+2h^2\right)\ln h+
h^2 - 4L_1 + 4\ln \Gamma_1\left(1+h\right)\bigg\rbrace\Bigg]\,,
\end{eqnarray}
where $h=\frac{H_{cr}}{2H}=\frac{m^2}{2eH}$, $L_1\cong 0.2487$,
$H_{cr}\cong4.4\times 10^{13}$ Gauss and $\Gamma_1(h)$ is the
generalized $\Gamma$ function. We observe that the expression in
the braces for the HEL contains terms with quadratic powers of $h$
as well as terms like $h\ln h$ and $h^2\ln h$. It is well known
that the logarithm of a trigonometric Cosine function has even
powers of the argument when expressed as a power series \cite{16,27,28}.
This is the situation encountered for $\Delta \mathcal{L}(h)$ when
the magnetic field H (also meant as B) is expressed as a cosine
wave for a plane electromagnetic wave propagation \cite{16}. The cases
of parallel and perpendicular polarizations for the weak and
strong field cases and the evaluation of the higher harmonics have
been treated in detail earlier in \cite{16}. Equation (28)
describes the case for arbitrary field strengths and generalizes
the evaluation for the generation of the higher harmonics.

\textbf{Magneto-optical Effect} We briefly discuss second and
higher-harmonic generation in a static magnetic field which is an
example of broken symmetry.  By broken symmetry, we mean a
violation of the principle of superposition/linearity. High
intensity synchrotron radiation, storage rings and lasers are good
experimental tools for the study of higher-harmonic generation.
The degeneracy of nonlinearity is broken by a static field or
nonplanar or nonmonochromatic waves and leads to the generation of
higher harmonics. We define,
\begin{eqnarray}
x = \frac{1}{\pi F_c} \,
\left[-S+(S^2+P^2)^{\frac{1}{2}}\right]^{\frac{1}{2}}\,, \quad
y = \frac{1}{\pi F_c} \,
\left[S+(S^2+P^2)^{\frac{1}{2}}\right]^{\frac{1}{2}}\,,
\end{eqnarray}
in analogy to a and b of equations (5) and (6) above. Here $F_c =
H_{cr}$. For
\begin{eqnarray}
x = 0, \quad y &\ne& 0, \quad  |P|=0, \quad
\mbox{we obtain} \nonumber\\[2ex]
{\rm Re} \, \mathcal{L}_{eff} &=&
S-\frac{\alpha}{\pi}\,F^2_c\sum_{k=1}\tilde{d}_{k}\,,
\end{eqnarray}
where,
\begin{eqnarray}
\tilde{d}_{k} =
-\frac{y^2}{2k^2}\left[e^\frac{k}{y}\,
{\rm Ei}\left(\frac{-k}{y}\right)
+ e^{-\frac{k}{y}}\,
{\rm Ei}\left(\frac{k}{y}\right)\right] = 
-\frac{y^2}{k^2}
\sum_{N=1}^{\infty}(2N-1)!\left(\frac{y}{k}\right)^{2N}\,.
\end{eqnarray}
Again, it is to be noted that Ei in Equation (31) is defined as real 
for real (negative and/or positive) arguments.

We use the formulas: $\zeta(4) = \frac{\pi^4}{90}$,
$\zeta(6)=\frac{\pi^6}{945}$ etc. 
to derive the equations above and below. $\tilde{d}_{k}$ can
be evaluated in the case of an external static magnetic field
$B_0$ along the x-axis. For an electromagnetic wave propagating
along the z axis, we have for the case \cite{7},
\begin{eqnarray}
E_x&=&E_z=0, \quad
B_y=B_z=0, \quad
|E_y|=|B_x|,\quad
P=E\cdot B=0,\\[2ex]
S &=& \frac{1}{2}\,
\left[E^2_y-(B_x+B_0)^2\right]=
-\frac{1}{2}\left(B^2_0+2B_xB_0\right)\,.
\end{eqnarray}
We then derive an expression for the partial derivative of the
${\rm Re}\mathcal{L}_{eff}$ with respect to S:
\begin{eqnarray}
\frac{\partial}{\partial S}\,
{\rm Re}\,\mathcal{L}_{eff}=
1+\frac{\alpha}{\pi}\left[\frac{4}{4S}\frac{S}{F^2_c}
+\frac{16}{105}\frac{S^2}{F^4_c}+
\frac{256}{315}\frac{S^3}{F^6_c}+\ldots \right]
\end{eqnarray}
The electric displacement field D can then be obtained.
\begin{eqnarray}
D=\frac{\partial}{\partial S}\, {\rm Re}\, \mathcal{L}_{eff}\cdot 
E\,.
\end{eqnarray}
We also observe that
\begin{eqnarray}
\frac{\partial P}{\partial E}\,
\cdot\,
\frac{\partial}{\partial P}\,
{\rm Re}\,\mathcal{L}_{eff}=0\,.
\end{eqnarray}
We consider electromagnetic waves with $E_y$ and $B_x$ of the
form:
\begin{eqnarray}
E_y &=& A\cos\phi=A\cos(wt-kz)\,,\\[2ex]
B_x &=& A\cos\phi=A\cos(wt-kz)\,.
\end{eqnarray}
We then find that,
\begin{eqnarray}
D_y & \cong &
E_y\left[1+\frac{4\alpha}{45\pi}\bigg\lbrace
-\left(\frac{B^2_0+2\,B_x\,B_0}{2F^2_c}\right)
+\frac{3}{7}\,
\left(\frac{B^2_0+2\,B_x\,B_0}{F^2_c}\right)^2
\bigg\rbrace\right]\,,
\end{eqnarray}
where $A$ is the amplitude of the wave. We find that the parallel
mode propagates alone. The $E_y B_x$ terms in equation (39)
indicate the presence of the second harmonic, as can be observed
from the fact that:
\begin{eqnarray}
2 \, E_y \, B_x = 2 \, A^2 \, \cos^2\phi = A^2 \, (1+\cos2\phi)\,.
\end{eqnarray}
The term proportional 
to $\cos(2\phi)$ indicates the presence of the second harmonic.
Similarly, $E_y B^2_x$ indicates the presence of the third harmonic
and $E_y B^3_x$ indicates the fourth harmonic etc. In a similar
way, the evaluation of $H$ can be done:
\begin{eqnarray}
H=-\frac{\partial S}{\partial B}\cdot \frac{\partial}{\partial S}\,
{\rm Re} \, \mathcal{L}_{eff}\left(P=0\right)\,.
\end{eqnarray}
The above expression of $H$ is also shown to exhibit the second
and higher harmonics. The spatial anisotropy is imposed by the
magnetic field. The SHG depends on propagation direction and
polarization direction of the fundamental electromagnetic 
wave with respect to
$B_0$. The maximum effect is when $\hat{B}_0$ is $\bot\hat{k}$
(the wave propagation direction).

%
%
\section{CONCLUSIONS}

We have investigated questions related to the representation of
the quantum electrodynamic (QED) effective Lagrangian and its
analytical expression. In Sect. 2, we briefly recalled our
previous results given for special-function representations of the
effective Lagrangian, and we briefly clarify the mathematical
notation used in the special-function representations (12) and
(21). The representation (21) unifies the real and imaginary
parts. We also introduced the Barnes function that should be of
use in a variety of applications that include the HEL. We very
briefly mention that the key step in the derivation of our 
series representations is the Mittag--Leffler theorem\cite{20}.

In section 3, we have also given an expression from the effective
Lagrangian that will facilitate the evaluation of the higher
harmonics. We also briefly discuss the magneto-optical effect of
the vacuum which might provide a signature of ``QED's nonlinear
light.'' In the Appendix, we discuss some aspects of the
renormalization and the renorm-invariance of the action.
Based on the results of the current paper, we
expect to carry out detailed studies related to various projected
and ongoing experiments and astrophysical phenomena [1,11--17]
involving strong static-field conditions (or fields with
frequencies that are small as compared to the electron Compton
wavelength). The study of the HEL to include finite temperature
effects \cite{29} with the series representation that we have developed
is an interesting problem that warrants further investigation.

%
%
\section*{ACKNOWLEDGEMENTS}

The authors wish to acknowledge insightful conversations with
Dr. Holger Gies and Professors Bowick and Joe Schecter of the
Physics Department, Syracuse University. They would also like to
thank Kato Lo and John Drozd for helpful discussions and 
invaluable assistance in processing the paper. 
S.R.V.~acknowledges a supporting grant from the Natural
Sciences and Engineering Research Council of Canada (NSERC). 

\appendix

\section{APPENDIX -- RENORMALIZATION\\
AND RENORMALIZATION--GROUP INVARIANCE}

The renormalization as well as the renormalization-group
invariance of the QED effective Lagrangian have already been
discussed at length by various authors (see \cite{30} and references
therein).

However, material on this topic is somewhat scattered in the
literature. In the current Appendix, we would like to give a brief
overview of the physical ideas that led to the formulation of the
related RG equations, as well as the connection with the $\beta$
functions of QED, complemented by an easy-to-understand
presentation of the basic notion of the renormalization itself.

It should be noted that the considerations in this section have
only partial significance for the main topic of the current
investigation, which is a treatment of higher-harmonic generation
based on series representations of the Lagrangian. The
current section merely provides illustrating remarks on the
derivation of the effective Lagrangian, which is again the {\em
starting point} of the main endeavour pursued.

The "original" Lagrangian of QED from which we start to work out
the $S$-matrix with all its radiative corrections in fact has to
be identified with the "bare" Lagrangian that is expressed in
terms of the "bare" physical quantities (bare charges, masses,
fields). As we develop the perturbations series (in $e^2$), we see
that some terms (it does not matter if they are infinite or not)
in fact modify the physical parameters that entered into the very
Lagrangian from which our work started. This means that we have to
renormalize the Lagrangian.

{\bf Renormalization:} The renormalization can be done by adding counterterms to the
Lagrangian. These terms relate the bare parameters (bare charges,
masses, fields) to the physical parameters of the theory, i.e. to
the renormalized charges, masses, fields. The renormalizability
then requires that the Renormalized Lagrangian $=$ Bare Lagrangian
+ Counterterms $=$ Bare Lagrangian but with charges, masses,
fields expressed in terms of the renormalized parameters. For QED,
we are in the lucky position that the theory remains finite in the
infrared, i.e. that the physical charge of the electron remains
finite as we move two electrons far apart. Therefore, we may
renormalize QED on mass shell, i.e. renormalize QED in such a way
that the renormalized charge is by definition the charge of an
electron as seen by another electron in the limit of a large
separation of the two electrons (this latter situation corresponds
to the limit of a very soft exchanged photon --- the infrared
limit in which the wave vector $k$ of the exchange photon tends to
zero).

To see how this works, look at p. 325 of \cite{31}: The vacuum
polarization modifies the Coulomb law (or Thompson scattering) in
such a way that ($e_0$ is the bare charge)
\begin{equation}
e_0^2/k^2 \to e_0^2/ [ k^2 ( 1 + \omega(k^2) ) ] = e_0^2/ [ k^2 (
1 + \omega(0) + O(k^2) ) ]\,.
\end{equation}
Now at $k=0$, the physical charge of the electron is obtained,
which means that $e^2 = e_0^2/(1 + \omega(0))$, or that the $Z_3$
renormalization that relates the bare and the renormalized charge of the
electron is obtained as $Z_3 = 1/(1 + \omega(0))$ where $e^2 = Z_3
e_0^2$. This relates the physical renormalized charge with the
bare charge of the electron. In order to construct a renormalized
Lagrangian in which the ``$e$'' is indeed the renormalized, finite
electron charge, we now have fulfill the condition that the
vacuum polarization as derived from our renormalized Lagrangian
results in a "modified" vacuum polarization correction to the
Coulomb law that fulfills $\omega_{\rm Renormalized}(0) = 0$. This
is just Eq. (8-96e) of \cite{31}. In general, a set of renormalization
conditions imposed on the renormalized Lagrangian determines the
physical interpretation of the parameters in terms of which the
renormalized Lagrangian is written. The renormalization conditions
(8-96) in \cite{31}, on which the considerations of \cite{20} are based,
ensure that the renormalized effective Lagrangian is written in
terms of the finite, renormalized, physical electron charge. The
"usual" representation of the QED Effective Lagrangian (QED
Effective Action) in Eq. (1) fulfills these conditions.

In other words, since we are used to defining the value of the
coupling by Thompson scattering, i.e., by scattering a
long-wavelength ($\omega\to 0$) photon off a (static) electron,
the renormalization scale is naturally set by the mass of the
electron. This amounts to the so-called $on-electron-mass-shell$
renormalization condition $e^2(\mu=m)/(4\pi)\simeq 1/137.036$. Our
starting point Eq.~(1) for the HEL is obtained exactly by
implementing this renormalization condition, and the coupling $e$
used throughout this work should be interpreted in this way. It is
precisely the second term in square brackets in Eq.~(1) that
guarantees this renormalization condition. This term subtracts any
contribution of $\Delta \mathcal{L}$ to Thompson scattering;
therefore, the latter is fully described only by the renormalized
Maxwell term together with the static-eletron interaction term
involving the on-shell renormalized coupling $e$ -- as it should.
This renormalization procedure has thereby fixed all free
parameters, such that the theory is now fully predictive for all
other processes that can be described by our one-loop HEL.

QED is usually defined at some high ultraviolet (UV) scale in
terms of a bare action. It is predictive for electrodynamic
processes at low energy scales to a high accuracy, once all
renormalization-group ``relevant'' parameters are specified. QED
has two such parameters, the coupling $e$ and the electron mass
$m$. During the derivation of the HEL (1), we have to give a
prescription how to fix these parameters to their physical values,
since they turn out to be scale-dependent. However, since the HEL
of Eq.~(1) describes processes with external photon lines to one
loop only, the scale dependence of $m$ remains invisible in this
calculation. Hence, to this order of accuracy, we can fix the
electron mass to its value measurable at the scale of, say, atomic
physics, and insert this into Eq.~(1). In the HEL language, the
scale dependence of the mass becomes visible at two-loop order,
see, e.g.,~\cite{30,32,33}. This subtlety emerges in the following way:
The unrenormalized two-loop expression contains a term which may
be written as $\delta m^2 \, (\partial/\partial m^2)$ (one-loop
Lagrangian), where m is the electron mass and $\delta m$ the
radiative modification (the physical mass is then $m^2_{\rm ph} =
m^2 + \delta m^2$. The original Maxwell Lagrangian is independent
of the electron mass. The one-loop Lagrangian that we work with
depends on the electron mass. The two-loop Lagrangian has
effective self-energy corrections to the dressed fermion
propagators that interact with themselves through a further
radiative photon. This self-energy effect is known to produce a
radiative renormalization $\delta m$ of the electron mass.
Therefore, the above term has to be interpreted as a correction to
the electron mass that enters into the one-loop Lagrangian,
expressing the fact that ``one-loop Lagrangian 
$+ \delta m^2 \, (\partial/\partial
m^2)$ (one-loop Lagrangian) $=$ one-loop Lagrangian as a function
of $m_{\rm Renormalized}^2 = m^2+\delta m^2$.'' This is the
mechanism by which the anomalous mass dimension creeps into the
renormalization group analysis of the action \cite{30}.

\textbf{Renormalization--Group Invariance:}
The QED Effective action is 
well known to be a
modification of the Maxwell Lagrangian relevant to the situation of 
strong background
fields. By contrast, vacuum polarization modifies the Coulomb at
small distances (large momenta). However, there is a connection
between the two regimes, asymptotically. Specifically, Ritus (see
\cite{32} and references therein) has shown that an interesting
connection exists between the corrections to the Maxwell equations
in intense fields and vacuum-polarization corrections to  QED at
large momenta. The connection can now be found upon considering
Eqs.~(25) and (26) of Ref.~\cite{32}. Observe that the mass
renormalization in Eq. (14) of Ritus also has the right asymptotic
behaviour at small proper time $s_0$. One of the basic ideas of
the renormalization group invariance is that the product of field
and charge, which enters into the covariant coupling, ${\rm i}
\partial_\mu - e_0 A^0_\mu \to {\rm i} \partial_\mu - e A_\mu$ in
going from bare to renormalized quantities, necessarily has to be
preserved under the renormalization because space itself
($\partial_\mu \equiv \partial/\partial x_\mu$) does not stretch
under the renormalization of charge and field. The charge and the
field are renormalized as $e_0 = Z_3^{-1/2} e$ and $A^0_\mu =
Z_3^{1/2} A_\mu$. [Here, ``0'' denotes the bare quantity.]

Let us finally discuss the scale dependence of the coupling $e$ as
it occurs in the HEL. Starting the computation with a bare
(Maxwell) action at UV scale $\Lambda$ with a bare coupling
$e_0\equiv e_0(\Lambda)$,  we obtain a $\Lambda$- and
$e_0$-dependent contribution to the Maxwell action $\sim
\mathcal{F}$. Together with the bare action, we can trade these
contributions for a scale-dependent coupling $e=e(\mu)$ that
describes the interaction of a photon with an electron at an
energy scale $\mu$~\cite{5,30,32}. Of course, $\mu$ is an arbitrary parameter,
and the requirement that no physical process should depend on
$\mu$ defines how the value of the coupling parameter $e(\mu)$ has
to be changed upon a variation of $\mu$. For quantitative
predictivity, we have to fix the value of $e(\mu)$ at one
particular scale $\mu$. Starting from this consideration, one may
derive functional relationships for the asymptotic behaviour of
the QED Effective Action in intense fields, which in turn enable
the direct transition to renormalization-group equations of the
Callan-Symanzik type [see Eq.~(45) of~\cite{32}]. Other types of
renormalization-group equations, for example those of the
Gell--Mann Low type, including a thorough discussion, can be found
in~\cite{5} and~\cite{30}.

Of course, the dependence on $\mu$ of the HEL could be formally
kept such as in~\cite{10} which would correspond to not specifying a
renormalization condition. However, once the input of a single
experiment (e.g., Thompson scattering) is taken into account, no
freedom is left in the parameters, the $\mu$-dependence has
disappeared and QED is fully predictive. In particular, there is
no room for further ``logarithmic correction terms''~\cite{10}.

\end{document}